\documentstyle[epsf,11pt]{article}
\def\nn{\nonumber}

\def \bi{\bibitem}

\def\d{{\rm d}}
 \def\(({\left(}
 \def\)){\right)}
\def\bi{\bibitem}

\def \ov{\over}

\def \b{\beta}

\def \d{{\rm d}}
\def \beq{\begin{equation}}
\def \eeq{\end{equation}}

\def \ov{\over}

\def \b{\beta}

\def \ab2{\alpha\beta^2}

\def \e{{\rm e}}
 \newcommand {\be} {\begin{equation}}
\newcommand {\bea} {\begin{eqnarray} \nonumber }
\newcommand {\ee} {\end{equation}}
\newcommand {\eea} {\end{eqnarray}}
 \newcommand {\eps} {\epsilon}

\input{epsf}
\begin{document}
\title{Constrained Boltzmann-Gibbs
measures and effective potential for glasses in hypernetted chain 
approximation and numerical simulations}

\author{Miguel Cardenas (*),  Silvio Franz(**) and Giorgio Parisi(***)\\
(*) Scuola Normale di Pisa\\
Piazza de Cavalieri 7,
56126 Pisa, Italy\\
(**) Abdus Salam International Center for Theoretical Physics\\
Strada Costiera 11,
P.O. Box 563,
34100 Trieste (Italy)\\
(***) Universit\`a di Roma ``La Sapienza''\\
Piazzale A. Moro 2, 00185 Rome (Italy)\\
e-mail: {\it cardenas@SABSNS.sns.it,
  franz@ictp.trieste.it, parisi@roma1.infn.it}
}
\date{January 1998}
\maketitle

\begin{abstract}
By means of an effective potential associated to a constrained equilibrium 
measure and apt to study frozen systems, we investigate glassy freezing 
in simple liquids in the hypernetted chain (HNC) approximation. 
Differently from other classical approximations of liquid theory,
freezing is naturally embedded in the HNC approximation. We get a 
detailed description of the freezing transition that
is analogous to the one got in a large class of mean-field long range 
spin glass. 
We compare our findings with Monte Carlo simulations of the same system
and conclude that many of the qualitative features of the 
transition are captured by the approximated theory. 
\end{abstract}
\vskip.5cm
\section{Introduction}
Cooled at fixed rate, supercooled liquids stop flowing on observable time scales when the glassy 
transition temperature $T_g$ is met \cite{vetro}.  At that cooling rate dependent temperature these 
systems get out of equilibrium.  The large scale motion of the molecules is frozen, and consequently 
the entropy (in the sense of the logarithm of the phase space accessible to the system) is 
discontinuously reduced.  This discontinuity is equal to the configurational entropy at temperature 
$T_g$ and is supposed to vanish if the system could be kept in equilibrium down to the point of the 
ideal glass transition $T_0$.  This would be observable only for infinitely slow cooling, and would 
be a real thermodynamic transition. 

At temperatures smaller then $T_g$ these systems finds themselves in a region of the configuration 
space having vanishing weight in the Boltzmann distribution\footnote{Strictly speaking the in the 
metastable region the liquid has zero weight, the Boltzmann distribution being concentrated on 
crystal configurations.  In this paper we simply neglect the existence of the crystal and we imagine 
that in the supercooled region the measure is concentrated on liquid configurations.  This situation 
can also be realized by not considering in the partition sum the crystal like configurations or by 
modifying the potential in such a way that the crystal get an high free energy.}, the values of the 
extensive quantities being far from these at equilibrium.  One can also expect that they will remain 
conformationally close to the configuration $y$ reached at $T_g$ where flow stopped.  On the other 
hand small scale vibrations of the atoms are free to thermalize at the actual temperature $T$ of the 
thermal bath.  This situation, with extreme separation of time scales would be naturally described 
by a statistical ensemble where the slow degrees of freedom are quenched, and the fast degrees of 
freedom thermalize at temperature $T$.  One can then define a conditional statistical ensemble, as a 
Boltzmann-Gibbs measure for fixed distance from the point $y$.  The free-energy associated to this 
distribution is a function of the constrained distance, and is a natural ``effective potential'' for 
glassy systems.  This effective potential can be computed with the replica method even in non 
disordered models, the role of the quenched variables being played 
 by the reference configuration 
$y$.  This reduces the problem to the study of the free-energy of a multicomponent mixture, in which 
an analytic continuation on the number of components has to be performed at the end of the 
computation.

In previous work \cite{I,letter,lungo}, it was studied
the shape of the potential, and the 
implications for the glass transition  in 
long-range spin glasses, 
and supported the generality 
of the picture in numerical simulations of a binary mixture model
\cite{lungo}. 
In this paper we extend the analysis to models of 
simple liquids in the HNC approximation, and show how 
this approximation, devised to study the liquid phase 
naturally describe glassy freezing \cite{mepa}. The same would not be true for 
other classical  approximations of liquid theory as the Percus-Yevick 
or the Mean Spherical approximations. The implementation of the replica 
formalism for the effective potential in the HNC approximation
is similar to the one used by Given and Stell \cite{given}
to study liquids in random quenched matrices. It also bear resemblance 
with the one used by Zippelius and coworkers to implement random 
crosslinking in models of vulcanization. The main difference is that, 
while in  these cases  the replica method is used to deal with 
external quenched disorder, in our case we use 
quenched degrees of freedom to probe the configuration space of 
systems that freeze even in absence of quenched disorder.

The approach of this paper is complementary to the one put forward 
recently in \cite{mepa}. There it was shown how, 
combining HNC and replicas, one could reveal the glassy transition and find the properties of the 
system below the glass temperature.  In this paper we will discuss the nature of the freezing in the 
HNC approximation finding a scenario very similar to that of mean-field spin glasses.  The 
approximation is constructed in such a way that the critical density 
automatically coincides with the one obtained in \cite{mepa}. The advantage of the ``effective 
potential'' framework with respect to that of
\cite{mepa} is to make conceptually clear the introduction of the replicas in the theory and to 
make testable predictions on the behaviour of the system at density less the critical one (as we 
shall see below) when we introduce a potential among two copies of the system.  On 
the other hand we will see that for high values of the density the simpler approach of this paper, 
where we neglect replica symmetry breaking, leads to inconsistency, and there one need to resort to 
the approach of \cite{mepa} for a coherent theory.

The explicit computations are performed  for the hard sphere potential.
We support our findings with Monte Carlo simulations of the same system. 
 A short account of our result has 
appeared in \cite{lett}.

We organize the paper as follows: in section 2 we discuss the 
construction of the effective potential and we briefly review the results 
obtained for long-range spin glasses. 
In section 3 we discuss the potential 
 for simple liquids in the HNC approximation. Section 
4 is devoted to the presentation of the theoretical results on the 
hard sphere system, that in section 5 we compare with the numerical 
simulations. Finally in section 6 we present some conclusions 
and perspectives.

\section{The effective potential}
In this section we review the construction of the 
effective potential \cite{I,letter,lungo}. 
For definiteness we discuss the case of a simple liquid 
composed by $N$ identical point-like particles in a volume $V$,
described by their coordinates $x=(x_1,...,x_N)$, and
 interacting
via a pair potential $\phi(x_i-x_j)$. 
Suppose that, undergoing a cooling process 
from the liquid phase, the system falls out of equilibrium at a temperature 
$T_g$ and remains stuck in a region of the configuration space having 
vanishing weight in the Boltzmann-Gibbs measure. This commonly happens at the 
glassy transition of supercooled liquids, where the liquid stops flowing:
large scale motion is frozen, while small scale motion 
of the atoms (vibration) can still equilibrate even below $T_g$. 
In these conditions the observed values of extensive quantities can 
be far from their canonical equilibrium  values, while keeping 
the external parameter constant, they do not vary over the laboratory
time scale. It is appropriate then to restrict the measure in 
configuration space to the vicinity of the configuration $y$ reached when 
crossing $T_g$. 

In order to do that we need to define a notion of similarity 
(or codistance) among configurations, $q(x,y)$, that, with reference to 
spin glass terminology, we call overlap. The appropriate definition 
of the overlap depends on the problem at hand, and it has to be such 
that to similar configurations correspond high values of $q$
(with normalization $q(x,x)=1$) and to very different configurations 
values close to zero. 

In our particle system  an appropriate definition can be 
\be 
q(x,y)={1\ov N} \sum_{i,j}w(|x_i-y_j|)
\ee
where $w(r)$ is a function close to 
one for $r\le \sigma r_0$ and close to zero 
for $r\ge \sigma r_0$, with $r_0$ being the radius of the particles and 
$\sigma$ a number e.g. of the order of $0.3$, such that
couples of particles at small distances in the two configurations 
contribute positively to $q$. Specifically in the unit radius 
hard sphere problem of section 4 
we will use $w(r)=\theta(r-0.3)$.
As $q(x,x)=1$ and $q(x,y)\le 1$ we can define a sort of distance as $d(x,y)=1-q(x,y)$.  In the 
following we will speak indifferently about the two quantities, remembering that high overlap means 
small distance and vice-versa.

Having now the definition of $q(x,y)$ we can define a restricted 
Boltzmann-Gibbs distribution 
as 
\be 
P(x|y)={1\ov Z(\b,y)} \exp\left(-\b H(x)\right) \delta(q(x,y)-q)
\label{p}
\ee
where $Z(\b,y)$ is the integral over $x$ of the numerator of (\ref{p}). 

Three comments are in order.
\begin{itemize}
\item
The value of $q$ that appears in (\ref{p}) is at this stage arbitrary.  However, the system at 
temperature $T$ will tend to adjust itself and select a given natural distance from the 
configuration $y$, according to local free-energy minimization.  The selection of $q$ can be well 
understood in a mean-field picture, and has been discussed in the case of long range spin glasses in
\cite{I}.  At low temperature one expects metastable states in configuration space.  This 
corresponds to a two-minima structure of the effective potential, with one minimum at low $q$ 
representing the typical overlap among configurations belonging to different metastable states, and 
one minimum at high $q$ representing the typical overlap among configurations in the same metastable 
state.  This last is the $q$ that would be naturally chosen by the system.

\item
The second comment concerns the dependence of the measure (\ref{p}) on the reference configuration 
$y$.  At a first sight, as different cooling experiment would produce different configuration $y$, 
it would appear that the measure (\ref{p}) could be hardly of any use.  However, $y$ is supposed to 
be a configuration {\it typical} with respect to the Boltzmann-Gibbs probability at temperature 
$T_g$, $\mu(y)=\exp(-\b_g H(y))/Z(\b_g)$, and we can expect the extensive quantities computed from 
(\ref{p}) to be self-averaging (i.e.  $y$ independent) in the thermodynamic limit.  The role of the 
configuration $y$ is analogous in this construction to the one of the quenched variables in 
disordered systems.  In this sense, the measure (\ref{p}) is an implementation of the idea of 
``self-generated disorder'' often advocated for structural glasses \cite{boume,kirtir}.  

\item
The third
comment concerns the selection of the temperature $T_g$.  For that we do not have any a-priori 
criterion, as it is a quantity that in experiments depends on the cooling rate.  We have thought to 
the temperature of the configuration $y$, that we will call $T'$ in the following, as the glass 
transition temperature in the purpose of illustrating the physical situation that we have in mind.  
As the matter of fact our construction is well defined for arbitrary $T'$, and interesting results 
are obtained even for $T'=T$.  In this paper we will limit out analysis to this case using the 
measure (\ref{p}) as powerful probe of configuration space.  We insist however on the conceptual 
importance of considering two temperatures, and we will often refer to results obtained in spin 
glasses for this more complicated case.  In the previous discussion and part of the following we 
have considered the temperature as the only external parameter.  It is clear that {\it mutatis 
mutandis} analogous considerations hold for any other control parameter, as the density of section 
5.
\end{itemize}

The object on which we will concentrate our attention 
is the free-energy associated to the distribution (\ref{p}) 
that, invoking the self-averaging property we can write as:
\be
V(q,\b,\b')=
- {T\ov N}{1\ov Z(\b') } \int \d y \exp(-\b' H(y))\log\left\{ 
 \int \d x \exp(-\b H(x)) \delta(q(x,y)-q)\right\}
\label{V}\ee
As the constraint implied by the delta function is global, 
we can enforce it through a Lagrange multiplier; and 
considering the quantity
\be
F(\eps,\b,\b')=
-{T\ov N}{1\ov Z(\b') }  \int \d y \exp(-\b' H(y))\log\left\{ 
 \int \d x \exp(-\b [H(x)-\eps q(x,y)]) \right\}
\label{F}
\ee
we find that $V$ and $F$ are related by the Legendre transform
\be 
V(q,\b,\b')=\min_{\eps}\left(F(\eps,\b,\b')+\eps q\right).
\ee
For practical purposes it is more convenient to 
work with $F$ than with $V$, while the data are more easily interpreted 
in terms of $V$. We will pass freely from one representation to 
the other in the following.

In order to deal with the average of the logarithm in (\ref{F})
we resort to the replica method. This consists in evaluating the 
moments $\overline{Z^r}$ for integer $r$ and computing average 
the logarithm 
from an analytic continuation to non integer $r$ from the 
formula $\overline{\log Z}=\lim_{r\to 0} {\overline{Z^r}-1\ov r}$.
Explicitly we can write: 
\be 
\overline{Z^r}=
-T \int \d x_0 \exp(-\b' H(x_0))/ Z(\b')
\int \d x_1...\d x_r  
\exp\left(-\b[\sum_{a=1}^r H(x_a)-\eps \sum_{a=1}^r q(x_a,x_0)]\right).
\ee
where we have written $y=x_0$.
The problem is reduced to the computation 
of thermodynamics for a mixture of $r+1$ components 
in the limit $r\to 0$. Notice the non-symmetric role played 
by the replica $x_0$ and the replicas $x_a$ for $a\ge 1$. This 
implies that while there is symmetry under permutation of replicas 
with positive index there is not symmetry under interchange of the replica 
$x_0$ with the others. Although we have $r+1$ replicas, the symmetry 
of the problem is only $S_r$, becoming $S_{r+1}$ only for 
$\eps\to 0$. 
Technically, our approach is similar to the one 
of Goldbart and Zippelius \cite{zipp} to study vulcanization of rubber
and the one of Stell et al. for liquids in random quenched matrices, 
where also one finds a number of replicas that tends to one.
However, in their case real quenched disorder is present, 
while in our case we use the auxiliary configuration $y$ to restrict the
Boltzmann-Gibbs measure to small regions of configuration 
space. 

Before discussing in the next section the application of the present 
formalism 
to simple liquids in the hypernetted chain approximation, let us 
describe briefly what one can expect for the effective potential 
when a glassy transition occurs, considering for simplicity the case 
$T'=T$. In the supercooled phase the diffusion 
constant becomes lower and lower  as the temperature is lowered.
The ``cage effect'' takes place: the 
molecules get trapped for long times before they can diffuse. 
When the glass transition is met diffusion is completely stopped,
at least on human time scale. The entropy associated to diffusion 
is lost at the transition.  Ergodicity is 
broken and the configuration space is effectively split 
into an exponentially large number of practically 
mutually inaccessible regions.

Making  the approximation that the 
time to jump out of these regions is infinite, 
 it is natural to expect the effective potential 
(\ref{pot}) to have two minima. One corresponding to the typical 
(low) overlap among configurations belonging to different regions
in  configuration 
space, and another corresponding to the typical overlap (high)
of different configurations belonging to the same region. 
The number of these regions, or metastable states ${\cal N}$ is related 
to the configurational entropy $\Sigma$ by the relation: 
${\cal N}=\exp(N\Sigma)$. The probability of $x$ to be in the same metastable 
state of $y$ will be in such conditions $1/{\cal N}=\exp(-N\Sigma)$.
Consequently the relative height of the high $q$ minimum with respect to 
the low $q$ one has to be equal to $T\Sigma$. 
This picture is realized and it has 
been discussed in \cite{I,letter} 
 in a large class of long range spin glass model.
The analysis of these models tells us that the picture has to be refined
a little. To each disconnected region one can associate a 
free-energy $f$, with an energetic part and an entropic part. 
Defining $\Sigma(T,f)$ the logarithm of the number of these regions 
as a function of $f$, one finds that, at low enough temperature, 
$\Sigma$ is different from zero in a finite temperature dependent 
interval $I(T)=[f_m(T),f_M(T)]$. The states that dominate the partition 
function
at temperature $T$ are such if the quantity 
\be
F=f-T\Sigma(T,f)
\label{dom}
\ee
is minimum \cite{pspin}. The study of the effective potential for 
$T\ne T'$ shows that individual states do not disappear 
when the temperature is changed, but they remain stable 
for large ranges of temperatures \cite{I,BBM,bfp}.
As the temperature is lowered, states with lower and lower $\Sigma$ 
are selected in  (\ref{dom}) until, for a temperature $T_s$ 
with  $\Sigma=0$ are reached and the partition starts to be 
dominated by the lowest states.

\section{The HNC approach}

Let us now start the discussion of the implementation 
of the HNC approximation in the  approach outlined in the 
previous section. As we stressed 
the use of the replica method reduces the problem 
of the evaluation of the effective potential to the one of an $r+1$
component liquid mixture, in which there is a privileged component 
with which all the other replicas interact via the potential
\be
-N\eps \sum_{a=1}^n q(x_a,x_0)=-\eps \sum_{a=1}^n\sum_{i,j}^{1,N}
w(x_i^0-x_j^a)
\ee

The basic quantities of the theory 
are the pair correlation functions among replicas 
\be
\rho_a\rho_b g_{ab}(x,y)+\rho_a \delta_{ab}\delta(x-y)
=\sum_{i,j}\delta(x_i^a-x)\delta(x_j^b-y)
\ee
A partial resummation of the Mayer expansion allows to write
a self-consistent expression for the free-energy \cite{61,macdo,mepa}:
\bea
-2\b F_{HNC} = & & \int \d^d x \sum_{a,b=0}^r \rho_a \rho_b g_{ab}(x)
\left[ \log g_{ab}(x) -1 +\b_a \phi(x)\delta_{ab}\right] \nonumber\\
 & &+2\b\eps\sum_{a=1}^r \rho_0 \rho_a g_{0a}(x)w(x)+{\rm Tr}\ {\bf L}(\rho h)
\label{fhnc}
\eea
where $h_{ab}=g_{ab}-1$ is the connected 
correlation function  and ${\bf L}$ is an  operator in 
physical and replica space,  
defined by 
\be 
{\bf L}(u)=u-u^2/2-\log(1+u).
\ee
We have also put $\b_0=\b'$,
$\rho_0=\rho'$ and $\b_a=\b$, $\rho_a=\rho$ for $a>1$. 
The free-energy (\ref{fhnc}) 
 has to be extremized with respect to the $g_{ab}$'s and the terms of order 
$r$ have to be extracted. 
The extremum conditions can be cast in  the form:
\be
g_{ab}(x,y)=\exp(-\b_a\phi(x,y)\delta_{ab} + \left\{\delta_{0a}(1-\delta_{0b})\b_b
+\delta_{0b}(1-\delta_{0a})\b_a\right\}\eps w(x)+h_{ab}(x)-c_{ab}(x))
\label{eqhnc}
\ee
with $g$ and $c$ related by the Ornstein-Zernike relation
\be
h_{ab}(x)=c_{ab}(x) +\sum_{c=0}^r\int \d y \  h_{ac}(x-y)\rho_c c_{cb}(y)
\label{oz}
\ee

As usual in the replica method one needs a parameterization
of the matrix $g_{ab}$ that allows the analytic continuation to $r\to 0$.
On the basis of symmetry considerations analogous to the 
one of  \cite{I}, one can propose the 
structure:
\be
g_{ab}=
\left\{
\begin{array}{cc}
g_{00} & a=b=0 \\
g_{10} & a=0,\ b\ne 0\ \ {\rm or}\ \ b=0,\ a\ne 0 \\
g^*_{ab} & a,b\ne 0
\end{array}
\right.
\ee
The $r\times r$
sub-matrix $g_{ab}^*$ can be either replica symmetric or 
have an ultrametric structure \cite{MPV}, in the following 
we will limit ourselves to the replica symmetric structure
 \be
g_{ab}^*=
\left\{
\begin{array}{cc}
g_{11} & a=b \\
g_{12} & a\ne b. 
\end{array}
\right.
\label{rs}
\ee
which coincides with the choice done by Given and Stell \cite{given}.
In the case of the p-spin model, the replica symmetric choice 
gives the correct result for the effective potential in the high and in the 
low $q$ regions and 
in particular around the minima. However it  was found an 
intermediate $q$  region where ``one step replica symmetry breaking'' 
in the matrix $g^*_{ab}$
was necessary to compute correctly the effective potential.
 Although
 we do not 
explore here the possibility of solutions with a structure more complicated 
then (\ref{rs}) and we limit ourself to the study of the high and low $q$
parts of the effective potential, we warn the reader that replica symmetry 
breaking has  also to be expected in this case for intermediate $q$. 

The interpretation of the different elements of the $g_{ab}$ matrix with
the replica symmetric  ansatz is straightforward. 
The element  $g_{00}$ represents 
the pair correlation function of the free system; as such the equation
determining it decouples from the other components in the limit $r\to0$. 
In turn, $g_{11}$ represents the pair correlation function 
of the coupled system.  $g_{10}$ is the pair correlation 
among the quenched configuration and the annealed one, while $g_{12}$
represents the correlation between two systems coupled with the 
same quenched system. This is the analogous of the Edwards-Anderson 
order parameter in disordered systems, and represents the long time 
limit of the time dependent autocorrelation function at equilibrium
\cite{I}. 

Straightforward algebra shows that the 
equations (\ref{oz}) reduce to the ones proposed in \cite{given}. 
In particular one finds that, as it should, the equation for $g_{00}$, 
describing the correlation function of the quenched replica decouples from the 
other and coincides with the usual HNC equation in absence of replicas. 
\bea 
h_{00}(x)&=&c_{00}(x)+ \int \d^d y \ 
[\rho_0 h_{00}(x-y) c_{00}(y)+ r\rho_1 h_{11}(x-y)
c_{11}(y)]\nn\\ 
h_{10}(x)&=&c_{10}(x)+ \int \d^d y \ 
[\rho_0 h_{00}(x-y) c_{10}(y)+ \rho_1 h_{10}(x-y)
(c_{11}(y)-c_{12}(y)(1-r))]\nn\\ 
h_{11}(x)&=&c_{11}(x)+ \int \d^d y \ 
[\rho_0 h_{10}(x-y) c_{10}(y)+ \rho_1 (h_{11}(x-y)
c_{11}(y)-h_{12}(x-y)c_{12}(y)(1-r))]\nn\\ 
h_{12}(x)&=&c_{12}(x)+ \int \d^d y \ 
[\rho_0 h_{10}(x-y) c_{10}(y)\nn
\\
&+& \rho_1 (h_{11}(x-y)
c_{12}(y)+h_{12}(x-y)c_{11}(y) -h_{12}(x-y)c_{12}(y)(2-r))]
\eea
 The overlap can be  expressed in terms of the correlation function 
$g_{10}$, and reads  
\be 
q=\rho \int \d x w(x)g_{10}(x)=4\pi \rho\int_0^\infty r^2 w(r)g_{10}(r).
\ee
According to the discussion of the previous section, we will associate 
glassy behavior to non convexity of the function $V(q)$, and in particular
to the existence of multiple  solutions $q(\eps)$ for $\eps\to 0$.
In the liquid phase we can expect instead $q(\eps)$ to be a single
 value function and a convex effective  potential with a minimum at 
$q=q_0=\rho \int \d x w(x)$, corresponding to the
absence of any structure in $g_{10}$, i.e. $g_{10}(x)=1$ for all $x$. 
A strong coupling $\eps$, attracting the system towards the configuration 
$y$ will force a structure in the $g_{10}$, which will have a higher peak 
in $x=0$ the higher is the coupling. Similarly $g_{12}$ will acquire a 
structure: if two different systems are similar to the configuration $y$ 
they will also be similar to each other. When the system freezes, it will 
exist a solution to the HNC equations in which $g_{10}$ and $g_{12}$
will have a structure even for small and vanishing $\eps$. 

The value of $\eps$ at which we find a solution with non-zero $g_{12}$ coincides with dynamical 
critical density of \cite{mepa} and it correspond to the phase transition point in a mode-coupling 
approach \cite{vetro}.  Indeed in this mean field approach ergodicity starts to be broken exactly at 
this point.

\section{Results for HNC hard spheres}

In this section we discuss the picture 
coming from the integration of the equations 
(\ref{eqhnc},\ref{oz}) in three dimension. 
We present systematic data in the case of the 
hard sphere potential 
\be
\phi(r)=
\left\{
\begin{array}{cc}
1 & r<1 \\
0 & r>1
\end{array} 
\right. 
{\ \ \rm and  \  \ with\ \ \ }
w(r)=
\left\{
\begin{array}{cc}
1 & r<0.3 \\
0 & r>0.3
\end{array} 
\right. 
\ee

\begin{figure}
\begin{center}
\epsfxsize=350pt
\epsffile{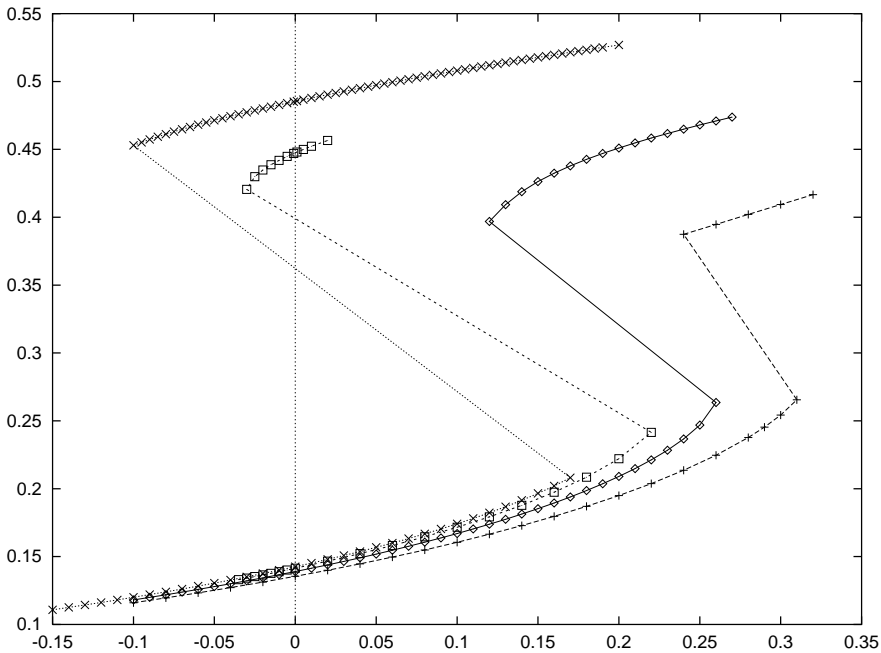} 
\end{center}
   \caption[0]{\protect\label{q-eps}  
The behavior of $q$ as a function of $\eps$ for 
HNC hard spheres for $\rho=1.14,1.17,1.19,1.20$. 
For high enough density $q$ is a multivalued function of $\eps$. 
We have shown only a portion of the curve in the region where it is multivalued.  For graphical 
transparency in this and the next figure we have joined with a line the branches corresponding to 
the same density.  } \end{figure} 

The hard sphere potential has been chosen for practical convenience, 
the glassy transition picture that will emerge can be strongly expected to 
be very general. We have verified in non systematic investigations 
that the same picture indeed holds for soft sphere systems with 
$\phi(r)=r^{-12}$. The value $0.3$ that appears in the definition of $w$
has obviously nothing fundamental, and we have checked that the picture 
is insensitive to its precise value. 
The hard sphere  model has no temperature, and the control 
parameter is the density.
Numerical work report a glassy phase for values of the density 
higher then 1.15. 

\begin{figure}
\begin{center}
\epsfxsize=350pt
\epsffile{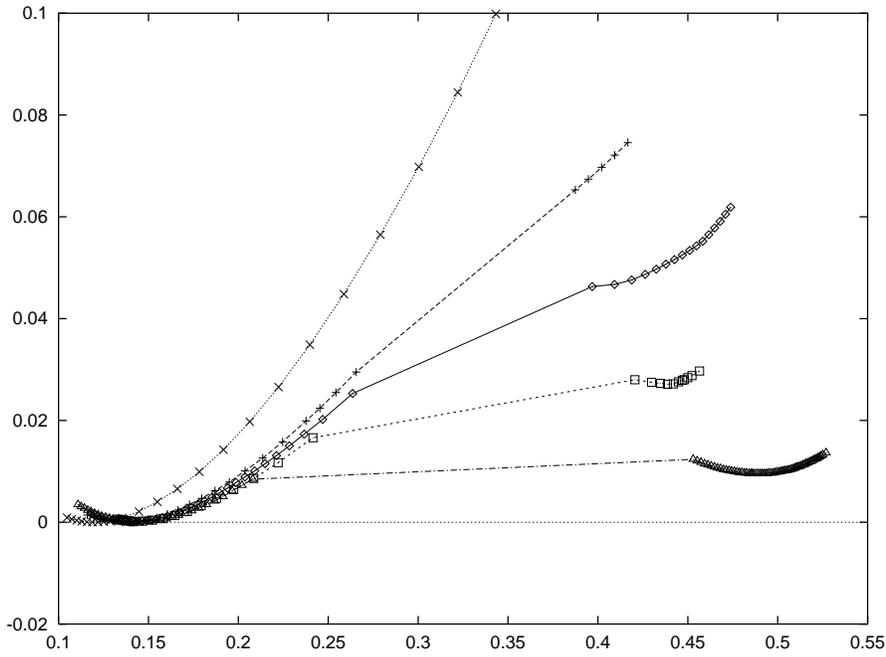} 
\end{center}
   \caption[0]{\protect\label{pot}  
The effective potential  for 
HNC hard spheres. From to to bottom $\rho=1.0,1.14,1.17,1.19,1.20$. For low 
density, high up in the liquid phase the potential is convex. 
In the glass phase two minima are present.} 
\end{figure}

We have solved 
 the saddle point equation (\ref{eqhnc},\ref{oz}) 
by iteration for various values of the density and the coupling. 
For fixed density we start the integration of the equation at low 
(respectively high) coupling $\eps$ 
where we know the solution and we increase (respectively decrease)
it at small steps. In this way we can find the curves of $q$ as a function 
of $\eps$, and reconstruct from (\ref{fhnc}) the effective potential $V(q)$.
We checked that up to a constant $V(q)=\int^q \d q' \eps(q')$. 
In the low density region we were able in this way 
to fully reconstruct the shape of the potential. For higher densities, 
we could just reconstruct in this way the high and the low parts 
of the effective potential. This is however enough to get a
fully detailed  picture of the freezing in the system. 
In figure \ref{q-eps} we present the curves of $q$ as a function 
of $\eps$ for various values of the density, while in figure \ref{pot}
we plot the corresponding curves $V(q)$.
 At low density $\eps$ is 
a monotonic function of $q$, testifying ergodic behavior of the 
system. The potential $V(q)$ is convex and has a single minimum 
for $\eps=0$, where the value of the overlap
is $q_0=\rho 4\pi (0.3)^3 /3=\rho\times 0.113$, corresponding to 
$g_{10}(x)=1$ for all $x$. 

\begin{figure}
\begin{center}
\epsfxsize=350pt
\epsffile{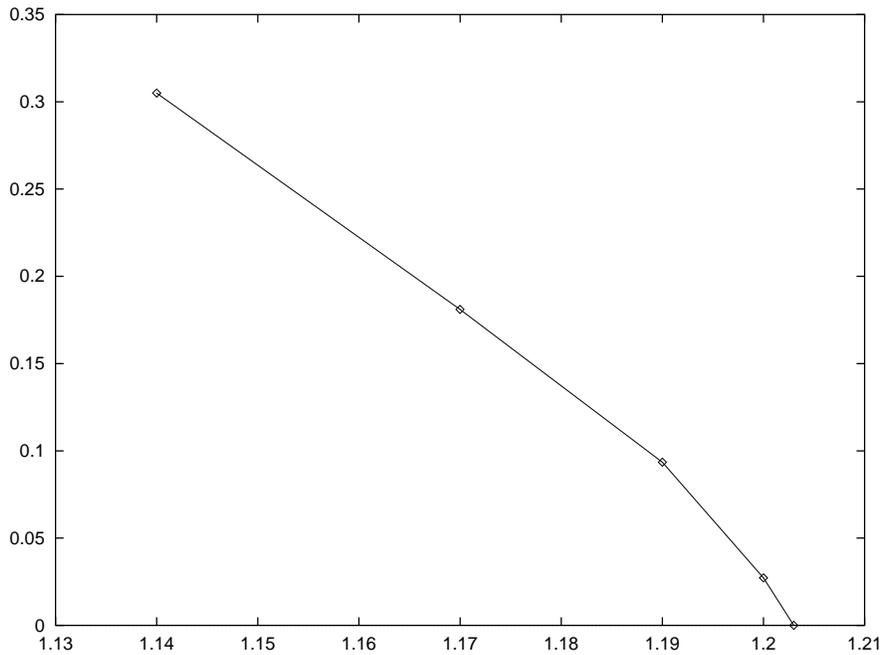} 
\end{center}
   \caption[0]{\protect\label{ph_dia}  
Phase diagram in the plane $\eps-\rho$. A first order transition 
line terminating in a critical point
separates a low $q$ from a high $q$ phase.}
\end{figure}

\begin{figure}
\begin{center}
\epsfxsize=350pt
\epsffile{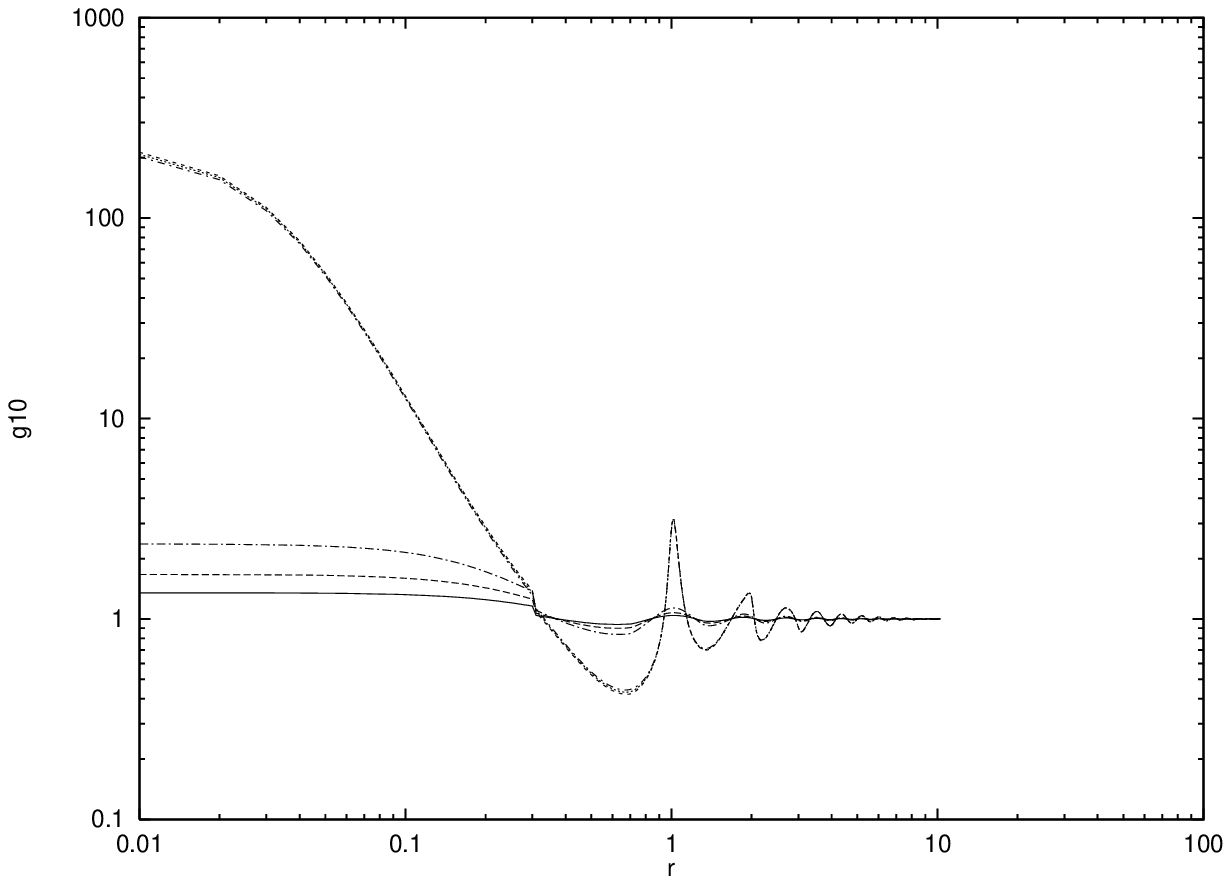} 
\end{center}
   \caption[0]{\protect\label{g10}  
The function $g_{10}(r)$ for 
$\rho=1.20$ and $\eps=0.10,0.15,0.20$. The three curves on the top correspond 
to the high $q$ solution, the three curves on the bottom to the low $q$ one.
 }
\end{figure}

Interesting behavior appears for densities higher or equal 
to $\rho_{cr}\approx 1.14$. At $\rho_{cr}$ the function $q(\eps)$
begins to be multivalued,  the potential looses the convexity
property and a phase transition among a low $q$ and a high $q$ phase can 
be induced by a coupling. The point $(\rho_{cr},\eps_{cr})$, with 
$\eps_{cr}=0.305$ is a critical point of second order phase transition, 
from which it departs a first order phase transition line $\eps_{tr}(T)$
(fig. \ref{ph_dia}).
The term $-N\eps q(x,y)$ in the Hamiltonian implies an energetic advantage
for the configurations $x$ close to $y$ and induces a transition 
between a high $q$ ``confined'' phase with high 
energy and a low $q$ ``deconfined'' 
phase with high entropy.
The transition line tells that generic equilibrium configurations
lie in metastable states for densities higher that $\rho_c$. 
For $\rho_{cr}<\rho<\rho_c=1.17$
the metastable states have a finite life, and a coupling 
$\eps\geq \eps_{tr}(T)$ is needed to stabilize them. At $\rho_c$
a minimum develops in the potential, and the 
metastable states have an infinite time life. 
The equation $q=q(\eps=0)$ has more then one solution. It has been 
shown by explicit calculation  in \cite{I} that a second minimum  
in the effective potential implies that in the equilibrium dynamics 
the system remains confined in a region with a large overlap with
the initial state.
In figure \ref{g10} we plot the function $g_{10}(r)$ for 
$\rho=1.20$ and various values of $\eps$ in the high $q$ and the low $q$ 
solutions. We see how the low $q$ solution has little structure ($g_{10}\approx
1$) while the high $q$ solution has a very pronounced peaks for integer 
values of $r$.

In ordinary cases  multiple minima in the effective potential as a 
function the order parameter signal the presence of different (stable 
or metastable) phases with different qualitative characteristics (e.g.
liquid and gas). Here the implications of the two minima structure are 
different. The appearance of the secondary minimum signals the breaking 
of the ergodicity, i.e. the split of the support of the Boltzmann-Gibbs 
measure into many, mutually inaccessible, regions. It is easy to realize
the link among  the two minima structure and  such non-ergodic situation.
If we suppose that the different region have typically all the same distance, 
one shall have  a minimum corresponding to that distance, and another 
minimum corresponding to the typical distance among configurations 
in the same region. In this perspective the minima are different manifestation 
of the same phase. Coherently with this picture, 
the internal energy in the two minima should be the same. 
Obviously this last sentence has no meaning  for hard spheres
 where the internal energy is not 
defined, but it can be easily checked in soft sphere systems. 
The difference of height 
between the two minima, $\Delta V$ can also be understood in this 
perspective as being due to the fact that the number of regions in 
which the configuration space has split (${\cal N}$)is
exponentially large in the number of particles ${\cal N}=\exp(N\Sigma)$, 
and each region carry a vanishing weight in the measure. 
In that conditions, chosen $y$ in a region
at random, the probability that $x$  falls in the same region, which 
should be $\exp(-\b \Delta V)$ is equal to
${\cal N}^{-1}=\exp(-N\Sigma)$. 

\begin{figure}
\begin{center}
\epsfxsize=350pt
\epsffile{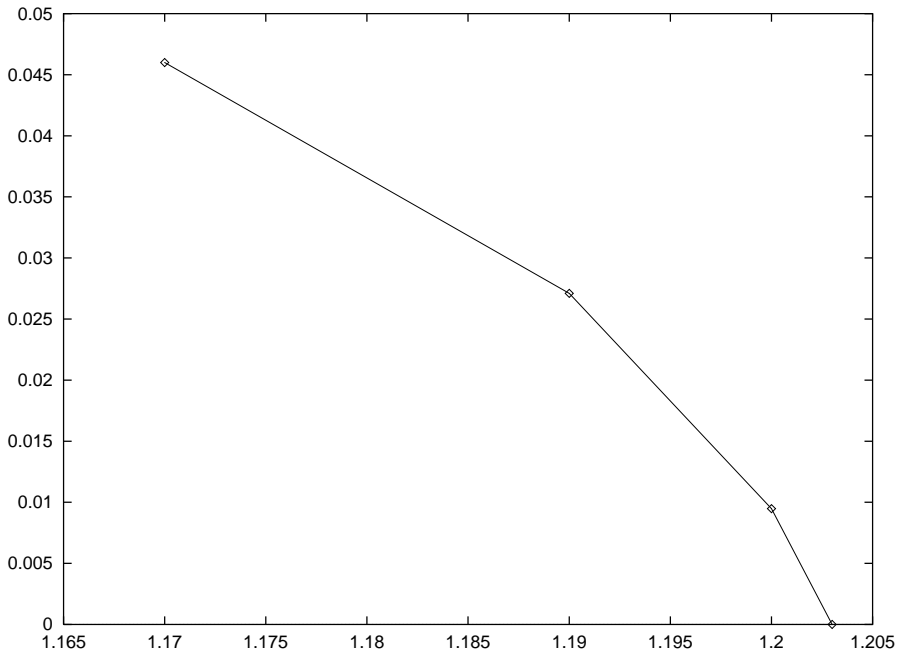} 
\end{center}
   \caption[0]{\protect\label{sigma}  
The configurational entropy $\Sigma$ as a function of $\rho$.} 
\end{figure}

We find then that $\Sigma$, to be identified with the configurational 
entropy, is related to $\Delta V$ by\footnote{the previous considerations
should be modified for $T'\ne T$ or $\rho'\ne \rho$. In that case 
the secondary minimum reflects the properties of the states of equilibrium 
at the primed values of the parameters when ``followed'' at the 
non primed values.} 
\be 
\Delta V=T\Sigma.
\ee
We have then a method to compute the configurational entropy, that 
we plot in figure \ref{sigma}. An equivalent method has been proposed in 
\cite{remi}. We see that $\Sigma$ is a decreasing function of the density 
and vanishes at a density $\rho_s\approx 1.203$,\footnote{The values of 
$\rho_c$ and $\rho_s$ are compatible with those found in \cite{mepa}, 
indeed the potential method reproduces the results of replica symmetry 
breaking for the static and the dynamic critical density.} 
according the scenario of Gibbs and 
Di Marzio of the glass transition, and analogously to long range 
spin glasses. At each value of the density one choose these states 
such that the total balance between $f$ the internal free-energy of the 
region and the configurational entropy is such to minimize the 
total free-energy.

Above $\rho_s$ the equations we are considering give 
the clearly unacceptable result of a negative configurational entropy, and the
approach must be modified. Previous experience in spin-glasses tells us that 
the paradoxical behavior has to be ascribed to an incorrect 
description of the quenched replica $y$. For $\rho >\rho_s$ 
the lowest $f$ states are chosen. These carry a finite Boltzmann
weight and a 
correct description has to take this into account. 
In the correct HNC approach to
 the high density regime the quenched replica should
 be  described by the replica formulation of M\'ezard and Parisi
with replica symmetry breaking \cite{mepa}. 
The HNC approximation, originally devised to study liquids at equilibrium, 
naturally embeds glassy behavior in a glassy transition scenario 
completely analogous to the one of disordered models with ``one step 
replica symmetry breaking''. 

\begin{figure}
\begin{center}
\epsfxsize=350pt
\epsffile{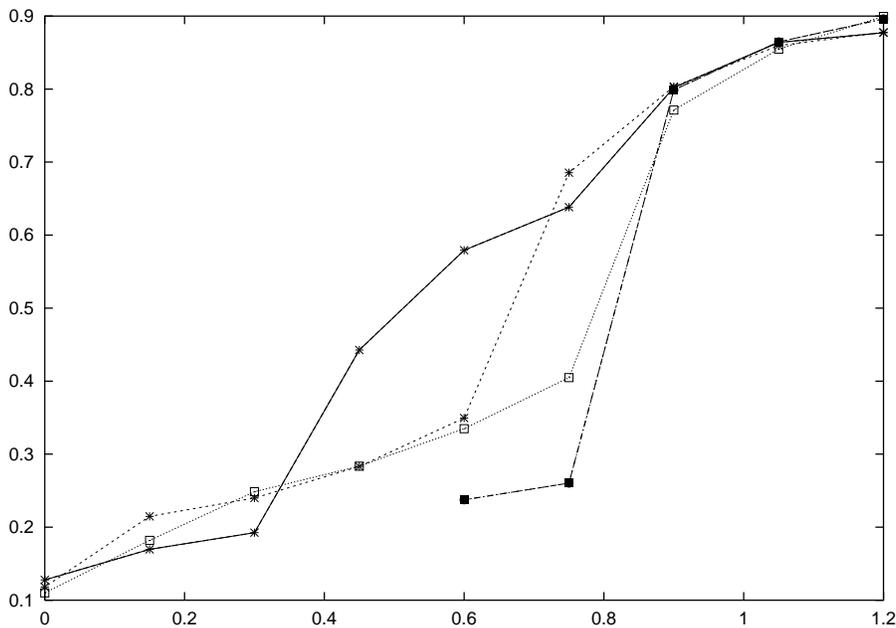} 
\end{center}
   \caption[0]{\protect\label{A}  
The behavior of $q$ as a function of $\eps$ for a system of 258 
particles and $\rho=1.04$. The different curves correspond 
different thermalization times $2^k$ for each value of $\eps$.
>From top
to bottom  $k=17,19,21,23$. For larger thermalization times the system
seem to develop a first order jump in $q$. 
 } \end{figure}
\begin{figure}
\begin{center}
\epsfxsize=350pt
\epsffile{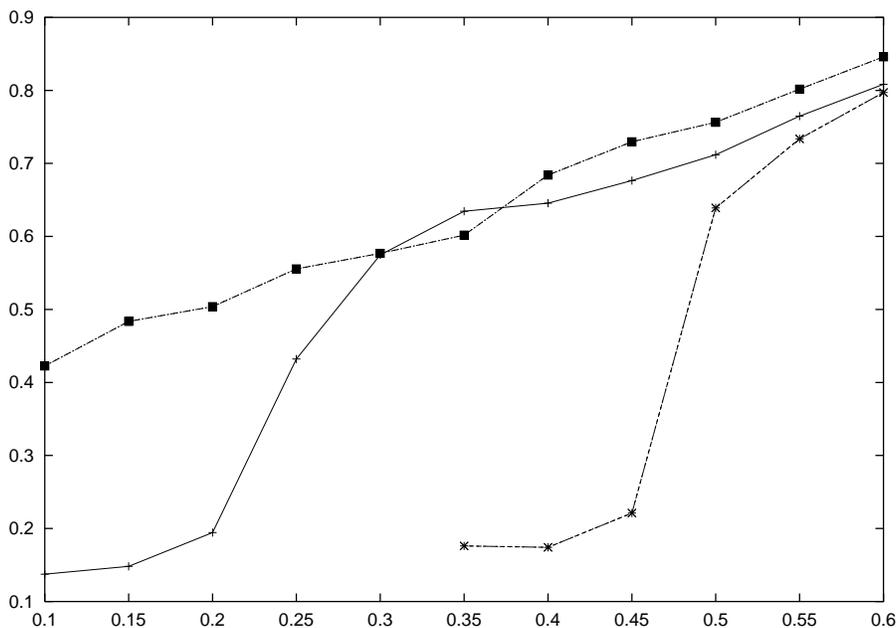} 
\end{center}
   \caption[0]{\protect\label{C}  
The same as figure \ref{A} but with an higher 
density, $\rho=1.10$ and  $N=1024$. The values of k are $k=17,19,21$. 
Notice that for $k=17$ the system has not relaxed to the low $q$ value 
even for $\eps=0.1$. 
 } \end{figure}

\begin{figure}
\begin{center}
\epsfxsize=350pt
\epsffile{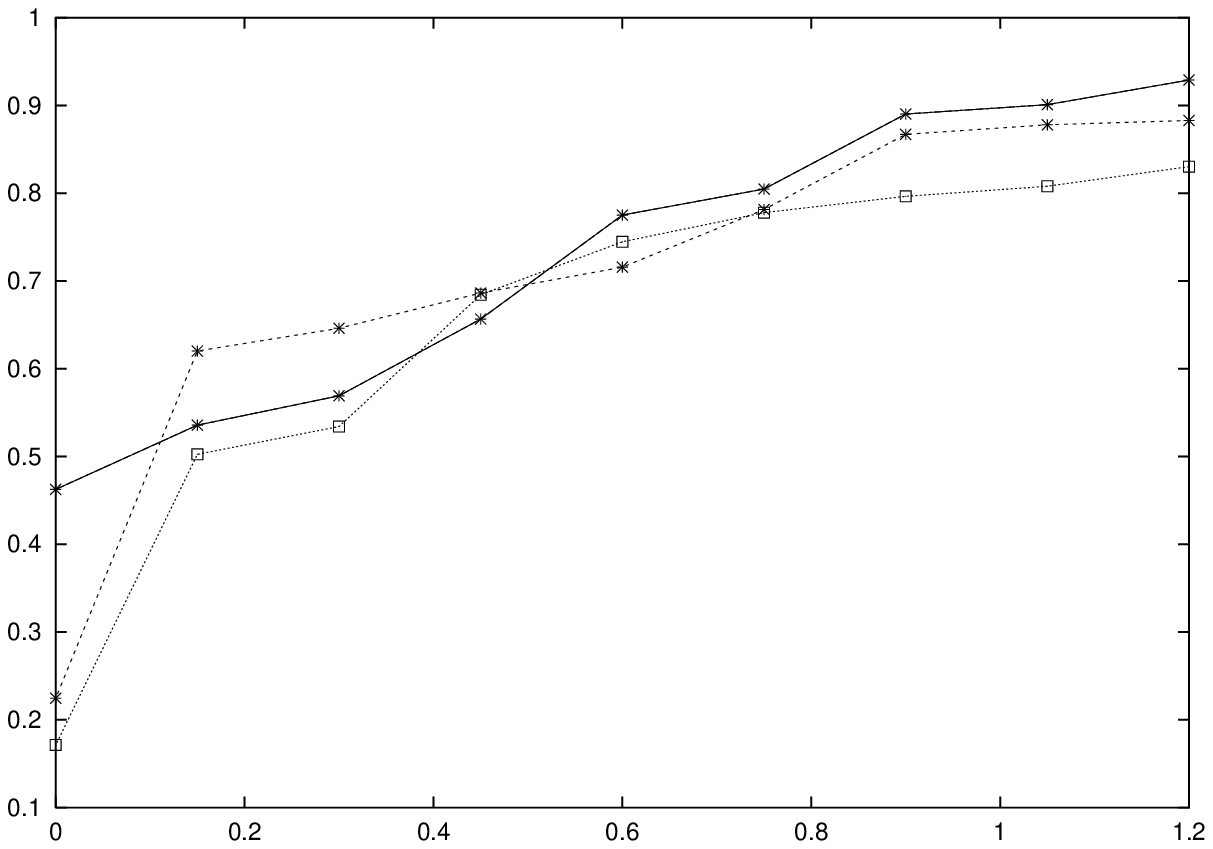} 
\end{center}
   \caption[0]{\protect\label{D}  
The same as figure \ref{A}   with $\rho=1.14$ and  $N=256$.
 The values of k are $k=17,19,21$. 
As expected, for
 higher densities the transition is pushed to lower values of $\eps$
 } \end{figure}

In many senses the picture is genuinely 
mean-field like. In fact, it has been stressed many times that 
in real systems metastable states with infinite life time do not 
exist, and a mechanism should restore ergodicity between $\rho_c$ 
and $\rho_s$. Moreover it is easy to realize that the effective 
potential we have defined must be convex beyond mean field. This can be seen 
constructing configurations with overlap inhomogeneous in space and 
with a free energy lower or equal to the convex envelope of the potential. 
We expect however a reflex of the mean-field structure 
in real systems. What it is seen as a sharp transition in mean field 
can still be observed as a crossover in real life and some of the 
prediction of mean field can be 
expected to hold in finite dimensional systems. 
In particular the existence of a first order transition line in the plane 
$\rho\ - \ \eps$, which depends on the existence of metastable states, 
regardless if their life is finite or infinite we expect to hold in real 
(or realistic) systems. In next section we will submit that to test. 

Before leaving this section let us remark that, 
how we show in some detail in the appendix, 
 the possibility of a 
glass transition, associated to nontrivial $g_{10}(x)$ for $\eps\to 0$ 
is excluded by other classical approximations 
of liquid theory, the Percus-Yevick approximation and the Mean Spherical 
Approximation. 

\section{Numerical Simulations}

In order to test the predictions of the transition scenario of the 
previous section we have performed Monte Carlo simulations of a system of hard spheres in three 
dimension coupled with a quenched configuration.  We have done simulations with a number of 
particles $N$ ranging from 256 to 1024 and we have not observed any significant dependence on the 
volume.

To generate the quenched equilibrium configurations at fixed density we start of $N$ particles of 
zero radius in a box with periodic boundary conditions, and we let the radii grow until two particle 
do not get in contact.  At that point we make a Monte Carlo sweep, i.e.  we move the particles of 
random amount and we accept the change if two spheres do not overlap; the size of the proposed 
move is fixed is such a way to have 0.4 average acceptance. We iterate the 
procedure until the desired density is reached.  The volume and the radius ($r$) are at the point 
rescaled in order to have $r=1$.  We thermalize then the system for 4000 Monte Carlo sweeps and use 
the configuration $y$ reached as external field for our coupled replicas experiment.  The relatively 
short thermalization is chosen in order to avoid crystallization.  We have careful verified that the
{\sl equilibrium} configuration does not have any signs of crystallization by monitoring his 
correlation function, which has a smooth minimum around 1.4, has it should do in the liquid phase.
 
Having generated the configuration $y$ we start the evolution of a coupled system $x$.  For various 
densities, we start the evolution from the configuration $y$ with an high value of $\eps$ and 
decrease the value of $\eps$ in units of $\delta\eps$, making $2^k$ Monte Carlo iterations for each 
value of $\eps$.  In figure
\ref{A}-\ref{D} we plot $q$ as a function of $\eps$ for various values of the density and different 
values of $k$.  We see that, as it should be expected for a system undergoing a first order phase 
transition, the curves are smooth for low $k$ and tend to develop a discontinuity for large $k$.  
For low density the discontinuity occurs at high $\eps$, while in is pushed toward small $\eps$ for 
high density.

\begin{figure}
\begin{center}
\epsfxsize=350pt
\epsffile{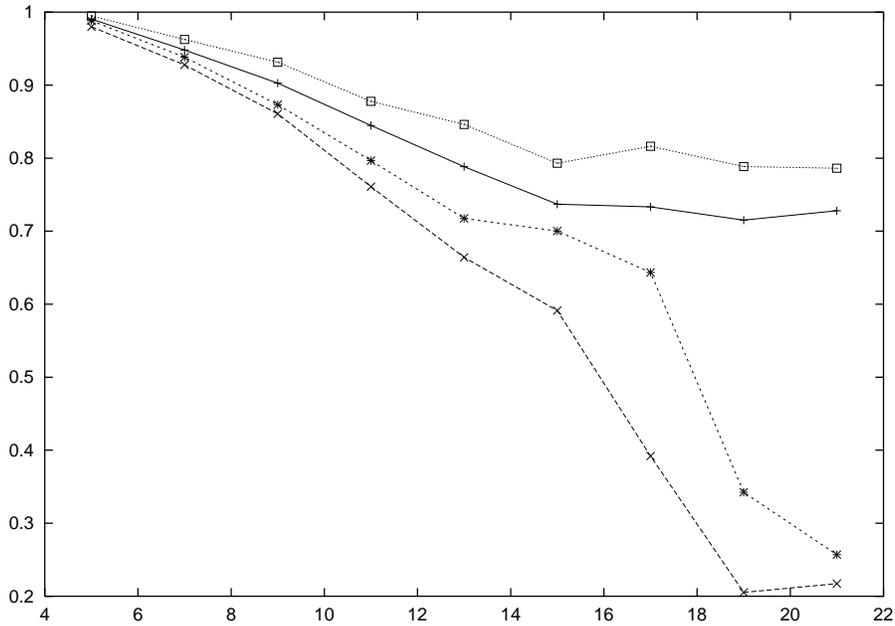} 
\end{center}
\caption[0]{\protect\label{E} The overlap as a function of logarithm (in base 2) of the time (i.e.  
the number of Monte Carlo sweeps)  starting from $y$ at time 0 and evolving for fixed $\eps$.  }
\end{figure}

A different numerical experiment is presented in figure \ref{E} where we plot the overlap as 
function of the number of Monte Carlo sweeps in a logarithmic scale.  Here we let the system evolve 
at fixed $\eps$ starting at time zero from $x=y$.  Again we observe a behavior compatible with a 
discontinuity of $q$ as a function of $\eps$.

\begin{figure}
\begin{center}
\epsfxsize=350pt
\epsffile{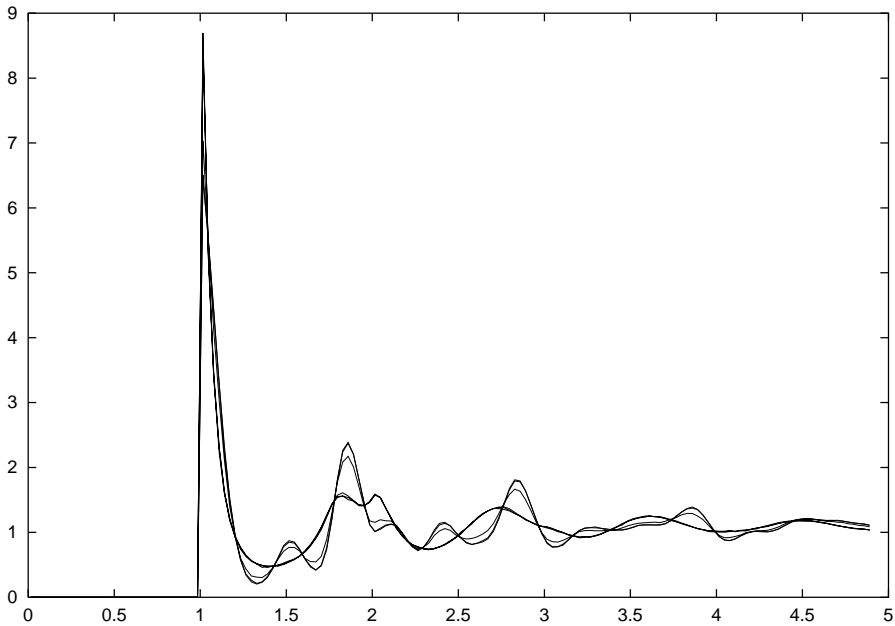} 
\end{center}
   \caption[0]{\protect\label{L}  
The function $g_{11}(r)$ for a system with $\rho=1.10$, $N=1024$ 
and coupling ranging from $0.35$ to $0.6$ in steps of $\delta \eps=0.05$
The curves do not variate too much with $\eps$, except the 
two with the lowest values of $\eps$. This could be a spurious effect 
do to partial crystallization as it is seen from the small peak 
around $r=1.4$. 
 } \end{figure}

\begin{figure}
\begin{center}
\epsfxsize=350pt
\epsffile{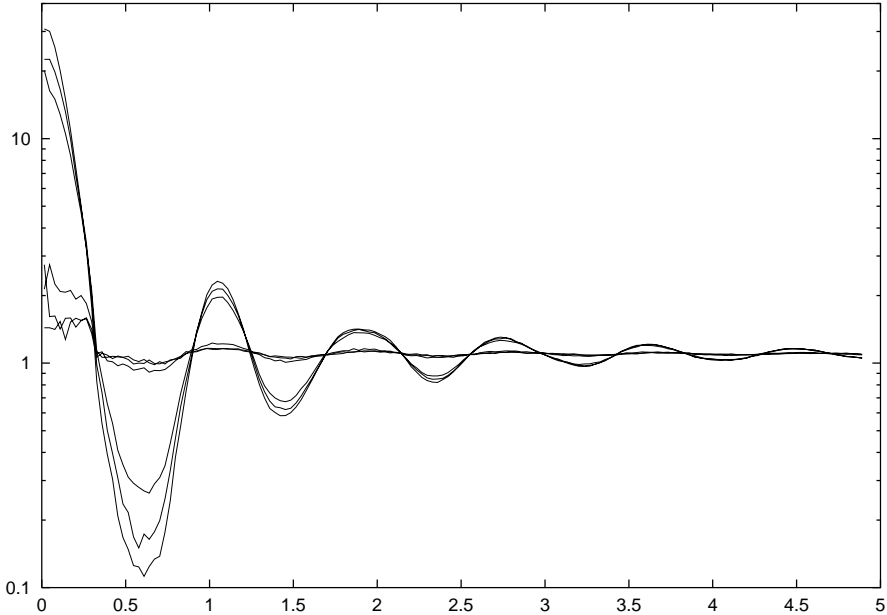} 
\end{center}
   \caption[0]{\protect\label{M}  
The function $g_{10}(r)$ for the same system of figure \ref{L}. 
We see a net discontinuity in behavior from the high to the low 
$\eps$ regions. 
 } \end{figure}

The same picture can be also observed directly in the behavior of the 
function $g_{10}(r)$. In figure \ref{L} and \ref{M} we can see that while 
$g_{11}$ does not vary too much as a function of $\eps$,\footnote{Notice the 
small peak at $r\approx\sqrt{2}$ for the curves with small $\eps$. This could 
signal some crystallization in the system.  In any case this effect is present only at high density 
and is absent in the simulations at lower density.  It seems that in the region where the overlap 
becomes quite small, the system crystallize in very long runs, while crystallization is obviously 
forbidden at higher values of the overlap.} one observes qualitatively different regimes in the high 
and in the low $\eps$ regime with a discontinuity in $g_{10}(0)$.  

For high $\eps$ we observe a 
strong oscillatory structure in $g_{01}$, which 
is very similar to $g_{11}$, with the only difference 
that the peaks are smoothed and by the presence of the large peak at $x=0$.  On the contrary in the 
low $\eps$ regime $g_{10}$ is very close to one for $r>0.3$ and has almost no structure.  Notice the 
striking similarity of this picture with the one that we get from the theory.

\section{Conclusions}
In this paper we have shown that constraining the Boltzmann-Gibbs measure 
to small regions of the configuration space we can study the transition 
from liquid to glass. We have studied in particular the case of the 
simple liquids in the HNC approximation. This model predicts a glassy  
transition in the lines of the Gibbs-Di Marzio scenario. 
The physical 
picture got from the present analysis is fully consistent with the 
one obtained from mean-field spin-glass models, although the HNC approach
allows to get information on the structure function of the supercooled 
liquid. The HNC approach, which consists in neglecting the so called 
bridge diagrams in the Mayer expansion turns in this way to be equivalent
to a kind of mean-field theory, where the glass transition is associated 
to a non-convex effective potential. Many of the prediction got from the 
theory are expected to be valid in real systems. In particular the picture 
of first order phase transition in presence of a coupling, which is 
associated to the presence of metastable regions in configuration space.
The picture has been positively  tested in numerical simulations 
on hard sphere systems. 

Growing evidence show that a good starting point for describing the glass transition of 
supercooled liquids is the scenario met in long-range spin glasses with ``one step replica symmetry 
breaking'' transition, which is the static counterpart of the schematic Mode Coupling Theory.  
Beyond these models this kind of transition pattern is met in mean-field glass models without 
disorder \cite{mapari,boume,fh}, and simulations of realistic glass models confirm many aspect of 
the picture \cite{pavetri}.  Finally the HNC approach of \cite{mepa} and with a different profile 
this paper also confirm this scenario.

We believe that the mean-field theory of the glass transition is now
on a firm ground. On one hand we have the mode coupling theory that 
allow to extract the dynamical behavior of supercooled liquids 
not too far from the glass transition. On the other, 
this theory has been repeatedly 
observed to be exact in mean-field spin glasses, whose 
study enrich the picture with a static view of the topography of the 
configuration space. In this paper we show that the same picture also
hold for the realistic model of liquid obtained by the HNC approximation. 

Finally we would like to comment once more on the limitation 
of the theory. As a genuine 
mean-field theory, the HNC predicts the existence of infinite life 
metastable states analogously to what happens in long range models 
and in the Mode Coupling approximation. According to this picture 
the system after having frozen into a metastable state at $T_g$, would 
``follow'' it down and back up in temperature in a completely 
reversible manner. 
While this can be true for short times, 
irreversible effects are observed on large time scales in glasses
\cite{struik}. The issue is related to the barrier jumping processes 
that restore ergodicity below $T_c$. 
Unfortunately at present the physics of these dynamical problem 
is unclear. 
 We hope that insight in  this problem 
will come from numerical and theoretical study of disordered models.

\section*{Acknowledgments} S.F. thanks the ``Dipartimento di Fisica dell' Universit\`a
di Roma La Sapienza'' for kind hospitality during the elaboration of this work.
   \section*{Appendix}

We show here that the Mean Spherical (MSA) and Percus-Yevick (PY)
approximations do not give any glass transition. The MSA deals 
with hard core interactions, for which $\phi(x)=\infty$ for 
$|x|<r_0$ and $\phi(x)=\phi_1(x)$ for $|x|>r_0$. In the canonical formalism 
the exact partition function can be written as a functional integral 
over the density field $n(x)$:
\bea
Z=& \int {\cal D}n(x) \ \exp
\left\{  
\int \d x 
\left[
-\b \phi(x-y) n(x)(n(y)-\delta(x-y))-n(x)\log n(x)
\right]
\right\}
\nonumber
\\
& \delta
\left(
\int \d x n(x)-\rho
\right)
\label{a1}
\eea
In the MSA one substitute the entropic part $-\int \d x n(x)\log n(x)$
in the exponent of (\ref{a1}) by the impenetrability constraint
$\int \d x n(x)n(x+y)=0$ if $|y|<r$ \cite{lp}. 
Expressing the constraint through 
a set of Lagrange multipliers $\lambda (x)$ one gets a Gaussian integral 
that can be computed exactly. 

In the formalism explained in the text the density variables
are replicated, and the constraint takes the form 
\be 
\int \d x \  n_a(x)n_a(x+y)=0 \;\;\;\;\;\; {\rm for} \;\; |y|<r_0 \;\;\;\;\;\; {\rm
and \;\; all}\;\; a
\label{a2}
\ee
Notice that the replica index in the two densities appearing in (\ref{a2})
is the same, as, obviously,
 there is no impenetrability constraint for different
replicas like there would be in a multicomponent fluid.  
Introducing now a set Lagrange multipliers $\lambda_a(x)$ to enforce
(\ref{a2}) and exploiting the properties of Gaussian integrals 
one founds that the correlation function 
\be
\langle 
n_a(x)n_b(y)
\rangle
-
\langle 
n_a(x)
\rangle
\langle 
n_b(y)
\rangle
=A_{ab}^{-1}(x-y)
\ee
can be found inverting the relation:
\be
A_{ab}(x)=\delta_{ab}\lambda_a(x)+V_{ab}(x)
\ee
where 
\be
V_{ab}(x)=\phi_1(x)\delta_{ab} - \{\delta_{0a}(1-\delta_{0b})\b_b
+\delta_{0b}(1-\delta_{0a})\b_a\}\eps w(x)
\ee
It is clear that for $\eps\to 0$, $A_{ab}$, and therefore its 
inverse, become diagonal in the replica indices, implying $g_{ab}(x)=1$
for all $x$ if $a\ne b$. 

Analogous reasonings apply to the PY approximation (see e.g. \cite{macdo})
where one has 
\be 
c_{ab}(x)=-g_{ab}(x)[\e^{\b V_{ab}(x)}-1].
\ee
For $\eps\to 0$ $c_{ab}$ becomes diagonal, and the Ornstein-Zernike 
relation (\ref{oz}) shows that  $g_{ab}(x)=1$ if $a\ne b$.

\end{document}